\documentclass[aps,preprint,nofootinbib,showpacs,superscriptaddress]{revtex4}
\usepackage{graphicx}
\usepackage{hyperref}
\usepackage{amsfonts}
\usepackage{amsmath,amssymb}
\usepackage{bm}
\usepackage{color}
\usepackage{epstopdf}
\usepackage{epsfig}
\usepackage{float}
{\rm }

\def\be{\begin{equation}}
 \def\ee{\end{equation}}
 \def\bea{\begin{eqnarray}}
 \def\eea{\end{eqnarray}}
\def\2{\frac{1}{2}}
\def\4{\frac{1}{4}}

\catcode`\@=11

\def\@normalsize{\@setsize\normalsize{15pt}\xiipt\@xiipt
\abovedisplayskip 14pt plus3pt minus3pt%
\belowdisplayskip \abovedisplayskip
\abovedisplayshortskip  \z@ plus3pt%
\belowdisplayshortskip  7pt plus3.5pt minus0pt}
\def\small{\@setsize\small{13.6pt}\xipt\@xipt
\abovedisplayskip 13pt plus3pt minus3pt%
\belowdisplayskip \abovedisplayskip
\abovedisplayshortskip  \z@ plus3pt%
\belowdisplayshortskip  7pt plus3.5pt minus0pt
\def\@listi{\parsep 4.5pt plus 2pt minus 1pt
            \itemsep \parsep
            \topsep 9pt plus 3pt minus 3pt}}

\def\underline#1{\relax\ifmmode\@@underline#1\else
        $\@@underline{\hbox{#1}}$\relax\fi}
\@twosidetrue \relax

\catcode`@=12

\catcode`\@=11

\def\section{\@startsection{section}{1}{\z@}{3.5ex plus 1ex minus
   .2ex}{2.3ex plus .2ex}{\large\bf}}
\def\ps@headings{\def\@oddfoot{}\def\@evenfoot{}
\def\@oddhead{\hbox{}\hfill
        \makebox[.5\textwidth]{\raggedright\ignorespaces --\thepage{}--
        \hfill }}
\def\@evenhead{\@oddhead}
\def\subsectionmark##1{\markboth{##1}{}}
}

\ps@headings

\catcode`\@=12

\begin{document}

\title{Duality between zeroes and poles in holographic systems with  massless fermions and a dipole coupling
}

\vspace{1.5cm}

\author{James Alsup}
\email{jalsup@umflint.edu}
\affiliation{Computer Science, Engineering and Physics Department,
The University of
Michigan-Flint,
Flint, MI 48502-1907, USA}
\author{Eleftherios Papantonopoulos}
\email{lpapa@central.ntua.gr}
\affiliation{
 Department of Physics, National Technical University of
Athens,
Zografou Campus GR 157 73, Athens, Greece}
\author{George Siopsis}
\email{siopsis@tennessee.edu}
\author{Kubra Yeter}
\email{kyeter@tennessee.edu}
\affiliation{
 Department of Physics and Astronomy, The
University of Tennessee, Knoxville, TN 37996 - 1200, USA
}

\date{\today}
\pacs{11.15.Ex, 11.25.Tq, 74.20.-z}

\vspace{3.5cm}

\begin{abstract}
We discuss the zeroes and poles of the determinant of the
retarded Green function ($\det G_R$) at zero frequency in a
holographic system of charged massless fermions interacting via a
dipole coupling. For large negative values of the dipole coupling
constant $p$, $\det G_R$ possesses only poles pointing to a Fermi
liquid phase. We show that a duality exists relating systems of
opposite $p$. This maps poles of $\det G_R$ at large negative $p$
to zeroes of $\det G_R$ at large positive $p$, indicating that the
latter corresponds to a Mott insulator phase. This duality
suggests that the properties of a Mott insulator can be studied by
mapping the system to a Fermi liquid. Finally, for small values of
$p$, $\det G_R$ contains both poles and zeroes (pseudo-gap phase).
\end{abstract}

\maketitle

\section{Introduction}

The AdS/CFT correspondence is a principle that connects a strongly
coupled $d$-dimensional conformal field theory with a weakly
coupled $(d+1)$-dimensional gravity theory
\cite{Maldacena:1997re}. This principle, also known as the
gauge/gravity duality, is well-founded in string theory, and has
been applied to many field theories having gravity duals with the
most noticeable application in condensed matter physics. One of
the earliest applications of the gauge/gravity duality to condensed matter
systems was the holographic description of a superconductor
\cite{Hartnoll}. It is  described by an Einstein-Maxwell-scalar
field theory in which the breaking of a gauge symmetry in the
gravity sector corresponds, in the boundary theory, to an operator
(the order parameter) which condenses below a critical
temperature signaling the onset of superconductivity.

The gauge/gravity duality has also been used as a framework for
Fermi liquid behavior \cite{Cubrovic:2009ye}, non-linear
hydrodynamics \cite{Bhattacharyya:2008jc}, quantum phase
transitions \cite{Iqbal:2010eh} and transport
\cite{Herzog:2007ij}. Thus, classical gravity theories have been
transformed into a laboratory for exploring condensed matter
physics from a different point of view \cite{Koutsoumbas:2009pa}.
In this way
 the holographic principle is a powerful tool
for studying strongly-correlated systems.

Recently, there has been considerable activity using the holographic
principle to describe metallic states of matter
outside of Landau's Fermi liquid theory
\cite{Liu:2009dm,Faulkner:2009wj}.  Experiments provide
evidence for the existence  of materials whose electronic
structure cannot be described by Fermi liquid theory, nor by any
other known effective theory. While Fermi liquids are
characterized by stable quasi-particles, these exotic materials
 possess well-defined Fermi surfaces but lack
long-lived quasi-particles.  A quantitative discussion of signatures of different
 excitations is presented in detail below.
To describe such metallic systems, one needs to introduce
charged fermions in the gravity sector of the holographic system.
However, in general, fermions cannot be treated classically as in
 the case of charged scalar fields of holographic superfluids or
superconductors. The coupled system of Einstein-Maxwell-Dirac
field equations become tractable only in the limit in which the
fermions may be treated locally in the bulk as a charged ideal fluid of
zero temperature. The solutions of these
systems are known as an electron star
\cite{Hartnoll:2009ns,Hartnoll:2010gu,Allais:2013lha}, Dirac hair
\cite{Cubrovic:2010bf} or confined Fermi liquid
\cite{Sachdev:2011ze}.

In another approach to describe the various
phases of a metallic state at low temperatures, a dipole coupling
for massless charged fermions  is introduced
\cite{Edalati:2010ww,Edalati:2010ge}.
The modified Dirac equation in the background of an
AdS-Reissner-Nordstr\"om black hole produces a
boundary fermionic spectrum with
vanishing spectral weight for a range of energies around $\omega=0$
without the breaking of any symmetry.
 As the dipole coupling strength is
varied, the fluid has Fermi,  marginal Fermi,  non-Fermi
liquid phases and also an additional insulating phase.  The
insulating phase possesses the characteristics representative
 of Mott insulators, namely
 the dynamic formation of a gap and spectral weight
transfer. Similar behavior was found in other constructions
\cite{Li:2011nz,Kuang:2012ud,Fang:2013ixa,Wu:2014rqa,PLH:2013}. The
dependence of the Fermi surface and selection of (non-) Fermi
liquid type are discussed in \cite{Guarrera:2011my}.

For a Fermi liquid, the characteristic signature of a
quasi-particle is a pole in the retarded Green Function $G_R$. The
location of the pole \be G_R^{-1} (\omega=0, k=k_F) = 0~, \ee in
momentum space at zero frequency dictates the Fermi surface and
dispersion. Additionally, the real component of the Green
function, $\Re G_R$, must change signs.  In a normal,
weakly-coupled metal, this is only possible through the pole, and
via Luttinger's theorem, integrating over the positive region of
$G_R$ one obtains the particle density \cite{Luttinger:1960zz}.

Apart from Fermi liquids, one of the principal examples of a
strongly correlated condensed matter system is the Mott insulator,
parent state of high-temperature copper-oxide superconductors in
which a gap is dynamically generated. An explanation for the
insulating nature \cite{SPC2007} is that a Mott gap manifests as a
surface of zeroes in the Green function.  The Green function of
interest, $G_R$, may be expressed in terms of the non-interacting
theory Green function, $G_0$, and the self-energy $\Sigma$,
\be\label{Gpole} G_R^{-1}(\omega, k) = G_0^{-1}(\omega, k) +
\Sigma(\omega, k)~, \ee of which a zero is indicative of a
divergent self-energy.  It is the divergent self-energy that
restricts the band from crossing the Fermi energy leading to
Mottness.  The full condition for Mottness is specifically on
the eigenvalues of the matrix $G_R$ and represented with the
determinant as,  \be\label{Mottness} \det
G_R(\omega=0,k=k_L) = 0~. \ee  For our purposes,
 we are interested in the spinor space but the statement may be
 extended to multiple bands \cite{Dzy2003}.  A mean-field theory approach
such as conventional BCS theory as calculated with the Nambu propagator,
 is unable to produce such a result.  The location of poles and zeroes in
 $G_R$ provides a path to uncover a system's behavior.

The interplay between poles and zeros is a characteristic feature
of QCD. The renormalization from high (UV) to low (IR) energy of
couplings and masses in QCD involves free quarks at UV scales,
which exhibit poles in the propagator, while at IR scales the
poles are converted to zeros \cite{'tHooft:1974hx} signaling the
formation of bound states.  The zero of the single quark propagator in the
IR is due to an orthogonal configuration between the (UV) single quark 
description and the composite mesons in the IR.

In condensed matter physics the
competition of poles ($k=k_F$) and zeroes ($k=k_L$) in the Green
function provides an illuminating criterion for characterizing the
different phases found in strongly-coupled fermion systems
\cite{Dzy2003}. Exploring the Hubbard model in this context has
led to a variety of features present in doped Mott insulators,
including spectral weight transfer, Fermi arcs found in the
pseudo-gap state, and Fermi pockets \cite{SPC2007,SKKRT2006}.
The Fermi arcs may be seen as a general consequence of a
composite state of UV fields, and hence zeroes, existing at low 
energies within the Hubbard model \cite{HP2012}.
In \cite{SMI2009}, a ``pole-zero mechanism'' for the two-dimensional
Hubbard model was used to characterize the transition from metal
to Mott insulator in terms of interfering pole and zero surfaces.
The formalism is comprehensive enough to accommodate arcs,
hole pockets, in-gap states, Lifshitz transitions and non-Fermi
liquids.  The work was furthered in \cite{SMI2010} to address
several anomalies in connection to (non) d-wave symmetries seen in
the pseudo-gap and to suggest a topological quantum phase
transition responsible for a non-Fermi liquid phase.

A holographic description of the structure of  poles and zeros
and their interconnection addresses central issues in strongly-coupled
fermionic systems.
A Fermi surface in the Green function for holographic fermions
was discovered in
\cite{Lee:2008xf,Liu:2009dm}
with the discussion further expanded
 in \cite{Faulkner:2009wj,Cubrovic:2009ye}. Near the
Fermi surface, the structure of the poles is expanded similarly to
eq.\ \eqref{Gpole} as \be\label{HGreen} G_R^{-1}(\omega, k) \sim
k-k_F -\omega/v_F - \Sigma(\omega, k) \ee with self-energy
$\Sigma(\omega, k)$ \cite{Faulkner:2010da} given as
\be\label{eqnupmT} \Sigma(\omega,k) \sim c
\omega^{2\nu_k^\pm}~~,~~~~6 \nu_k^2 = k^2  - \frac{q^2}{2}~, \ee
with fermion charge $q$ and numerically calculated constant
$c$.  It is this scaling dimension $\nu_k$ that
selects the leading term in (\ref{HGreen}) dictating the
dispersion relation at the Fermi surface, and hence the type of
fluid.
 For scaling dimension $\nu_k < 1/2$, the pole of $G_R$
 corresponds to an unstable quasi-particle.
  This is identified as a non-Fermi fluid.  With the value
$\nu_k =1/2$, the excitations are of marginal Fermi fluid type.
For $\nu_k > 1/2$ the dispersion relation is linear.  This is the
Fermi (non-Landau) fluid. Lastly, imaginary $\nu_k$ corresponds to
``log oscillatory'' solutions.

Since the discovery of Fermi surfaces within the strongly coupled
field theories analyzed via holography, an exploration of fermions
has taken place.
Mechanisms underlying different p-wave \cite{pWave}
and d-wave \cite{BHY2010}  symmetries for  spectral
functions have been offered.  An effective lattice was added in
\cite{LSSZ2012} by using a modulated chemical potential. The
spatial dependence alters the energy scaling and
offers identifying signs for a holographic realization of a pseudo-gap
state.

More fundamental (top-down) holographic
theories from  truncations of string/M-theory have produced viable models
describing superfluity and superconductivity
\cite{Gubser:2009qm,Gauntlett:2009zw,Gauntlett:2009dn,Gauntlett:2009bh,Bobev}.
In higher dimensions,
fermions are typically coupled to gravity and gauge fields with
high-order derivatives.
There are consistent truncations to lower dimensions in
the fermionic sector giving interesting fermionic couplings
\cite{Bah:2010yt,Bah:2010cu}.
Dipole (Pauli) couplings are found to be a common feature of the theories
with the coupling constant realized as scalar-dependent \cite{DGR:2012}.
It would be interesting to explore
their effect on fermion spectral functions at the boundary.

In this work, we aim at  studying the zeroes and poles of the Green
function $G_R$  in a holographic system with a gravity sector
consisting of an AdS-Reissner-Nordstr\"om black hole and a dipole
coupling of  massless charged fermions to an electromagnetic field.
 By exploring a duality
between the zeroes and poles of this holographic system we show
that by varying the dipole coupling strength, $G_R$ may possess
only poles, only zeroes, or both poles and zeroes.

This behavior suggests that the dipole coupling strength
plays the role of an order parameter in the Mott physics. By
studying a simple model in which a fermion is coupled to a gauge
field through a dipole interaction in the bulk, we find that on
the field theory side  as the strength of this interaction is
varied, a new band in the density of states emerges and spectral
density is transferred between bands. In the language of Green
functions only poles is the signature of the Fermi and non-Fermi
liquids, only zeroes corresponds to the Mott insulating phase, and
the coexistence of both poles and zeroes is the pseudo-gap phase.

This  duality between zeroes and poles also suggests that
within the holographic system, the properties of the Green
function $G_R$ signifying the Mott phase can be inferred by
mapping to the well understood Green function of a Fermi liquid 
by simply changing the sign of the dipole coupling
constant $p\to-p$.

The discussion is organized as follows. In Section (\ref{sect2}) we set
up the theory and derive the Einstein-Maxwell-Dirac equations.
In Section (\ref{sect3}) we solve the Dirac equation. In Section
(\ref{sect4}) we discuss the duality of poles and zeros, and
finally,  Section (\ref{sect5}) contains our conclusions.

\section{Gravitational Bulk}
\label{sect2}

 The bulk dynamics in an asymptotically Anti-de
Sitter (AdS) space is described by the Einstein-Maxwell action
with cosmological constant $\Lambda = -3/L^2$, \be S = \int d^4 x
\sqrt{-g} \left[ \frac{R + 6/L^2}{16\pi G} - \frac{1}{4} F_{MN}
F^{MN}\right]~, \ee where $F_{MN} =\partial_M A_N - \partial_N
A_M$ is the field strength of the $U(1)$ vector potential $A_M$.
For convenience, we set $L=4 \pi G = 1$.

The Einstein-Maxwell equations admit a charged four-dimensional
AdS black hole solution, \be ds^2 = \frac{1}{z^2} \left[ - h(z)
dt^2 + \frac{dz^2}{h(z)} + dx^2 + dy^2 \right]~. \label{eqmetric1}
\ee The metric function is given by
 \be\label{eq6}  h(z) = 1 -\left( 1 +\mu^2 \right) z^3 + \mu^2 z^4 ~,\ee
 with the horizon radius set at
$z=1$, and the $U(1)$ potential is \be\label{eq3} A_{t} = \mu
\left( 1 - z \right) \ , \ \ A_z =A_x=A_y =0~, \ee corresponding
to a non-vanishing electric field in the radial $z$ direction,
\be\label{eq5} F_{tz} = - F_{zt} = \mu~. \ee The Hawking
temperature is given by \be\label{eq7} T = - \frac{h'(1)}{4\pi} =
\frac{3-\mu^2}{4\pi }~, \ee with $\mu^2 = 3$ providing the zero
temperature limit. There is a scaling symmetry of the solutions
found as \be  z \ \to \lambda z \ , \ \ x\to \lambda x \ ,  \ \
\mu \ \to \mu /\lambda \ , \ \ T \to T/\lambda~, \ee and we should
only report on scale-invariant quantities, such as $T/\mu$,  etc.

In this gravitational background we add a massless fermion with
charge $q$ and we include a dipole coupling to the $U(1)$ field
\cite{Edalati:2010ww,Edalati:2010ge}. The action is \be
S_{\mathrm{fermion}} = i\int d^4 x \sqrt{-g} \bar{\Psi} \left[
/\hspace{-.25cm}D  - p \Sigma^{MN} F_{MN} \right] \Psi ~. \ee
 The various terms in the action are \bea
/\hspace{-.25cm}D &=& e_a^M \Gamma^a \left(\partial_M + \Omega_M - i q A_M \right) ~,\nonumber\\
\Omega_M &=& \frac{1}{8} \omega_{abM} \left[ \Gamma^a , \Gamma^b \right]~,\nonumber\\
\omega_{abM} &=& \eta_{ac} \omega^{c}_{bM}~,\nonumber\\
\omega^a_{bI} &=& e^a_M \partial_I e_b^M + e^a_M e_b^N \Gamma^M_{NI}~,\nonumber\\
\Sigma^{MN} &=& \frac{i}{4} \left[ \Gamma^a ,\Gamma^b\right] e_a^Me_b^N ~,
\eea
with spin connection $\omega_{abM}$ and vierbein $e_a^M$, and lower-case indices $a,b$ belong to the tangent space.

In the conventional case, $p=0$, the system is of two
non-interacting Weyl fermions of  opposite chiralities. With $p\ne
0$, the dipole term introduces an interaction between
the two Weyl fermions. The system corresponds to an order
parameter in the dual gauge theory of conformal dimension $\Delta
= \frac{3}{2}$.

To analyze the Dirac equations of the bulk fermion we find it more
convenient to go to momentum space by Fourier transforming
\be\label{diracEqn}
 \Psi =  e^{-i\omega t - i k x } \sqrt{\frac{z^3}{h}} \left(\begin{array}{c} \psi_{-} \\ \psi_{+}
 \end{array} \right)~~,~~~~\psi_\pm = \left(\begin{array}{c} \psi_{\pm1} \\ \psi_{\pm 2} \end{array} \right)~,
\ee and  choosing a basis for  the $\Gamma$ matrices \bea \Gamma^1
= \left(\begin{array}{cc} i \sigma_1 & 0 \\ 0  & i \sigma_1
\end{array} \right) ~~&,&~~~~\Gamma^2
= \left(\begin{array}{cc} - \sigma_2 & 0 \\ 0  & - \sigma_2 \end{array} \right)~,\nonumber\\
\Gamma^3 = \left(\begin{array}{cc} 0 & -i \sigma_2 \\ i \sigma_2
& 0 \end{array} \right) ~~&,&~~~~\Gamma^4 =
\left(\begin{array}{cc} - \sigma_3 & 0 \\ 0  & - \sigma_3
\end{array} \right)~. \eea The Dirac equation decomposes into
decoupled equations \bea
\pm h\psi'_{-,12}   + \left(\mu q (1-z) + \omega \pm k \sqrt{h} \mp p\mu \sqrt{h}\right) \psi_{-,21} &=& 0,\nonumber\\
\pm h\psi'_{+,12}  + \left(\mu q (1-z) + \omega \mp k \sqrt{h} \mp p\mu \sqrt{h}\right) \psi_{+,21}
 &=& 0~.\nonumber
\eea The presence of the dipole coupling $p$ modifies
the Dirac equation. The effects of the  coupling will be
seen more clearly in the solutions of the Dirac equations.

\section{Solutions of the Dirac Equations}
\label{sect3}

To solve the Dirac equations we choose in-going boundary
conditions at the horizon,

\be \psi_{\pm,12} = \left(1-z\right)^{-i\omega /(4\pi T)} \mathcal
F_{\pm,12}~. \ee The ratios \be \xi_\pm = \frac{\mathcal F_{\pm
1}}{\mathcal F_{\pm 2}}~, \ee satisfy the non-linear flow
equations~\cite{Faulkner:2009wj} \be\label{FlowEqn}
 h \xi'_\pm + \left[\omega + \mu q (1-z) + \sqrt{h} \left(\pm k+ \mu  p z \right)\right]  \xi_\pm^2  + \omega + \mu q (1-z) + \sqrt{h} \left(\mp k-\mu p z\right) =0~,
\ee
together with the in-going boundary conditions,
\be
\xi_\pm = \left\{ \begin{array}{ccc} i &, & \omega\ne 0~, \\ \\
 i \sqrt{\frac{  q/\sqrt{2}+\sqrt{3} p\pm k
  }{
q/\sqrt{2} - \sqrt{3} p \mp k  }} & , & \omega =0~. \end{array}
\right. \ee The solution to the flow equations determines the
retarded Green function as \be G_R(\omega,k) =
\left(\begin{array}{cc} G_+ & 0 \\ 0 & G_- \end{array} \right)~, \
\ G_\pm (\omega,k) =  \left. \xi_\pm\right|_{z\to 0}~.
\label{gsolution} \ee
%and spectral function by $A(\omega,k) = \frac{1}{\pi} \Im Tr G_R(\omega,k)$.
%
From the symmetries of the Dirac equation \eqref{diracEqn}, we
deduce the relation between the Green functions \be\label{Gsym}
G_\pm (\omega,k) =   G_\mp(\omega,-k)~. \ee The Fermi momentum is
found as the pole \be\label{eqFe} G_\pm^{-1} (\omega = 0,k=k_F) =
0~. \ee Since both solutions $\psi_\pm$ are normalizable, there is
an alternative quantization in which $G_+$ and $G_-$ are
interchanged.\footnote{Since the appearance of this work,
the duality between poles and zeroes
has subsequently been shown as a consequence of the two possible
quantizations \cite{VP:2014}.}
This leads to physically equivalent results.

To identify the type of (non-) Fermi fluid, we look at the
near-horizon region, $z\to 1$.  Employing the  limiting procedure
\cite{Faulkner:2009wj}, we obtain the scaling dimension
$\nu_k^\pm$ given by \be\label{eqnupm} 6 \left(\nu_k^\pm \right)^2
= \left( \sqrt{3} p \pm k \right)^2  - \frac{q^2}{2} ~, \ee
to be compared with (\ref{eqnupmT}).
 The
self energy $\Sigma(\omega , k)$ at the Fermi surface remains
 \be
\Sigma(\omega,k) \sim c \omega^{2\nu_k}~. \ee
For our system with fixed fermion
charge and mass, $\nu_k$ and the fluid character is determined by
the strength of the dipole interaction $p$.

\begin{figure}[t]
\includegraphics[width=.4\textwidth]{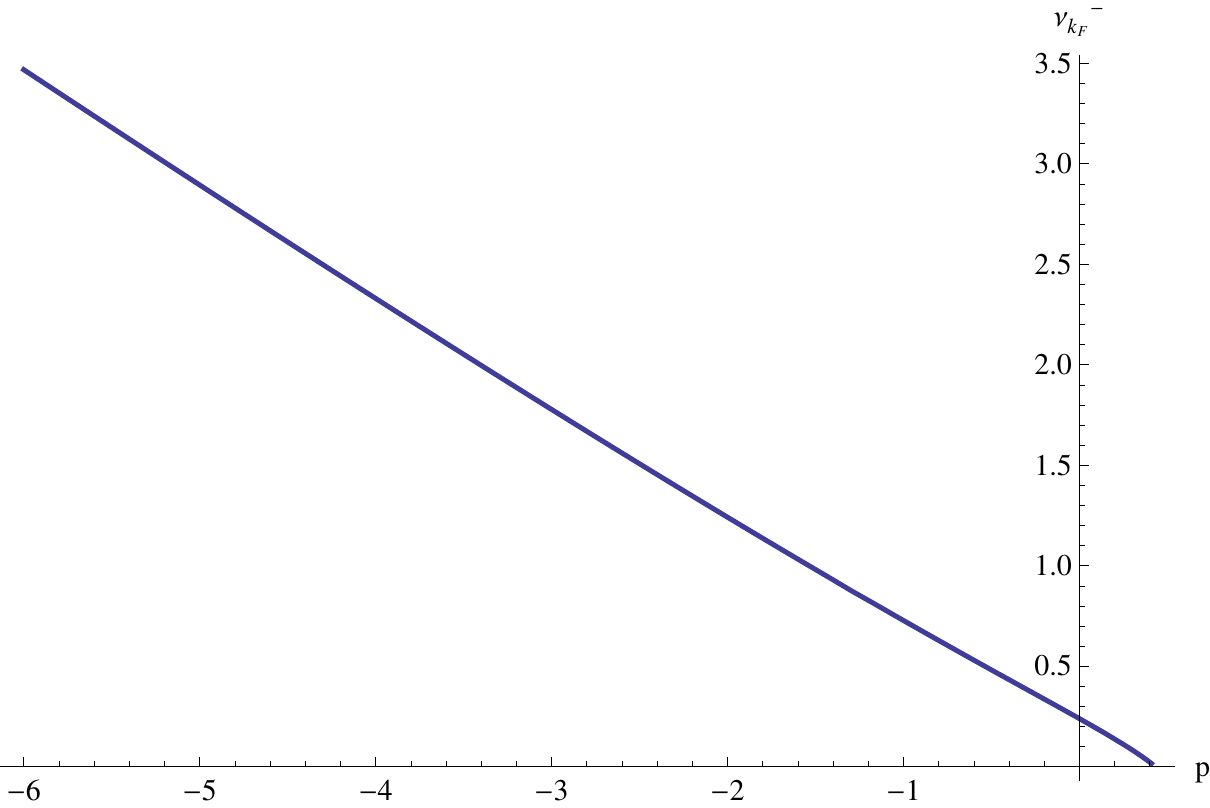}
\caption{Scaling dimension $\nu_{k_F}$ \emph{vs.}~$p$ for $q=1$.}
\label{knuVsp}
\end{figure}

To study explicit solutions, we set the fermion charge to $q=1$,
and numerically solve the flow equation (\ref{FlowEqn}) to
determine where the system  possesses a Fermi surface and the type
of excitations.  Figure \ref{knuVsp} displays our results for
$\nu_k^\pm$ evaluated at the Fermi momentum $k_F$.
We recover the results of \cite{Edalati:2010ww,Edalati:2010ge} and
classify the excitations as Fermi  liquid ($p \lesssim -.53$),
marginal Fermi liquid ($p \sim -.53$), non-Fermi liquid ($ - .53
\lesssim p \lesssim .41 $) and log-oscillatory ($p\gtrsim .41$).
The system has a Fermi surface with $p \lesssim .41$ but no Fermi
surface for larger values of $p$.

\section{Duality of poles and zeroes}
\label{sect4}

Using the properties of the solutions
(\ref{gsolution}) of the flow equation (\ref{FlowEqn}) we will
show that there is a duality between poles and zeroes.
We first emphasize that a pole of $G_\pm$ at $\omega =0$
(eq.~\eqref{eqFe}) is not necessarily a pole of the determinant
\be \det G_R = G_+ G_- ~.\ee It is known \cite{Faulkner:2010da}
that in the conventional case, $p=0$, \be\label{eqknown} \det G_R
(\omega =0 , k; p=0) =1~, \ee therefore it possesses neither poles
nor zeroes. This is because poles (zeroes) of $G_+$ are  cancelled
by zeroes (poles) of $G_-$ at the same momentum. We shall see that
this coincidence of poles and zeroes is lifted when the dipole
coupling is turned on, resulting in poles and zeroes of $\det
G_R$.

\begin{figure}[t]
\includegraphics[width=.4\textwidth]{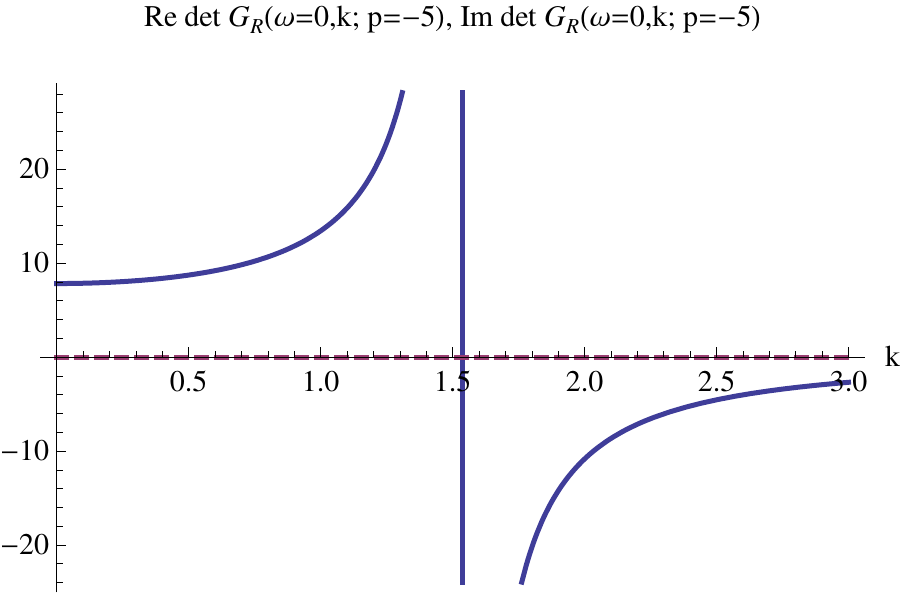}
\includegraphics[width=.4\textwidth]{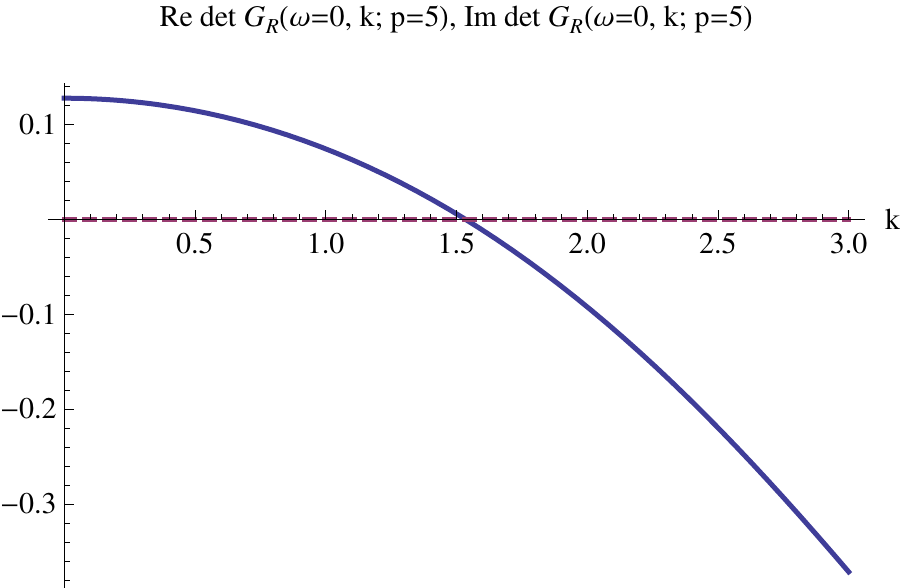}
\caption{Plots of $\Re\det  G_R$  ($\Im \det G_R =0$) with $q=1$ for $p=-5$ (first panel)
showing a pole at $k=k_F \approx 1.5$, and $p=5$ (second panel) showing a zero at $k=k_L \approx 1.5$.}
\label{detReG}
\end{figure}

It is easily deduced from the flow equation \eqref{FlowEqn}, that
the reciprocal of $\xi_\pm$, \be \zeta_\pm = \frac{1}{\xi_\pm}~,
\ee
 satisfies the flow equation
\be
 h \zeta'_\pm  + \left[-\omega - \mu q (1-z) + \sqrt{h} \left(\pm k+ \mu  p z \right)\right] \zeta_\pm^2
 - \mu q (1-z) + \sqrt{h} \left(\mp k- \mu p z\right) - \omega
=0~. \ee Upon comparison of the two flow equations, we see that
$\zeta_\pm$ solves the same equation as $-\xi_\pm$ under the
change of parameters $k\to -k$ and $p\to -p$. It follows that the
inverse Green function $G_\pm^{-1}(0,k)$ at $p$ is identified with
$-G_\pm(0,-k)$  at opposite dipole coupling $ -p$.  Using the
relation between the two Green functions \eqref{Gsym}, we arrive
at
\be\label{HoloDet} \det G_R (\omega=0,k;p) = \frac{1}{\det G_R
(\omega=0,k;-p)}~. \ee This is the central result of our work and
its importance relies on the fact that it establishes   a relation
between poles and zeroes of  the determinant of the Green function
at zero frequency between systems of opposite dipole coupling. For
$p=0$, we recover eq.~\eqref{eqknown}. For $p\ne 0$, the poles
found for $p<0$ (Fermi liquid) correspond to zeroes for $p>0$
(Mott insulator).

\begin{figure}[t]
\includegraphics[width=.4\textwidth]{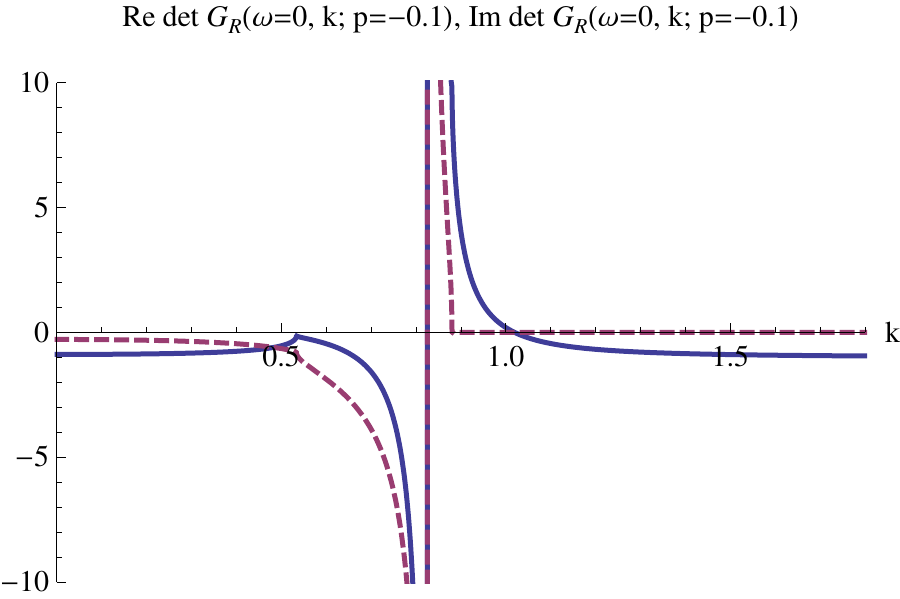}
\includegraphics[width=.4\textwidth]{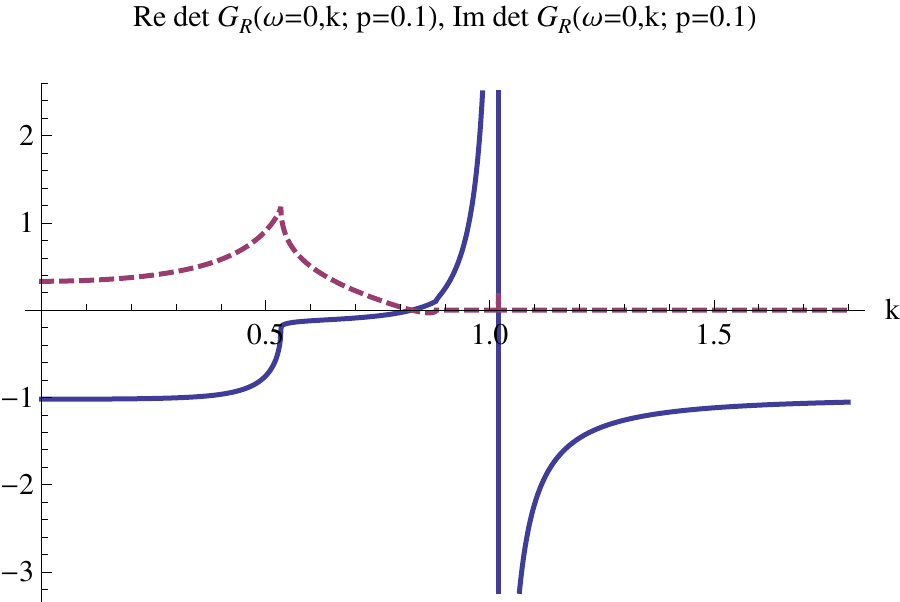}
\caption{Plots of $ \Re \det G_R$ (solid lines) and $ \Im \det
G_R$  (dashed lines)  with $q=1$ for $p=-0.1$ (first panel)
showing a pole at $k=k_F \approx 0.8$ and a zero at $k=k_L\approx
1.0$, and $p=0.1$ (second panel) showing a pole at $k=k_F \approx
1.0$ and a zero at $k=k_L \approx 0.8$.} \label{detGp0p1}
\end{figure}

For large negative dipole coupling strength, i.e., $p \lesssim
-.53$, we are  in the Fermi liquid phase, since $\nu_k > 1/2$
(figure \ref{knuVsp}). In this regime, we obtain poles of $\det
G_R$ and no zeroes. A typical example is shown in figure
\ref{detReG} (first panel) for $p=-5$. In this case, $\Im \det
G_R$ vanishes and the graph has two poles at $k=k_F\approx \pm
1.5$, and no zeroes. According to the duality \eqref{HoloDet}, we
expect to see two zeroes at $k=k_L \approx \pm 1.5$ and no poles
for a system with $p=5$. This is indeed what we obtain by a
numerical calculation of $\det G_R$ (second panel of figure
\ref{detReG}). This is in the Mott insulator regime and no Fermi
surface is found.

For small values of the dipole coupling strength ($|p| \lesssim
.41$),  we expect both zeroes and poles. Again, using the duality
\eqref{HoloDet}, the zeroes can be deduced from the poles at
opposite $p$ (and vice versa). An example is shown in figure
\ref{detGp0p1} for $p=\pm .1$. Unlike in the case of large $p$,
$\Im \det G_R$ does not vanish. Instead, the zeroes for $p=.1$
found at $k=k_L\approx \pm .8$ are isolated zeroes of both
$\Re\det G_R$ and $\Im \det G_R$. Correspondingly, the same is
true for the poles at $k=k_F \approx \pm .8$ for $p=-.1$. At the
other set of zeroes at $k=k_L \approx \pm 1.0$ for $p=-.1$ (and
correspondingly poles at $k=k_F \approx \pm 1.0$ for $p=.1$), $\Im
\det G_R$ vanishes over a range containing the zeroes (poles).

\begin{figure}[t]
\includegraphics[width=.45\textwidth]{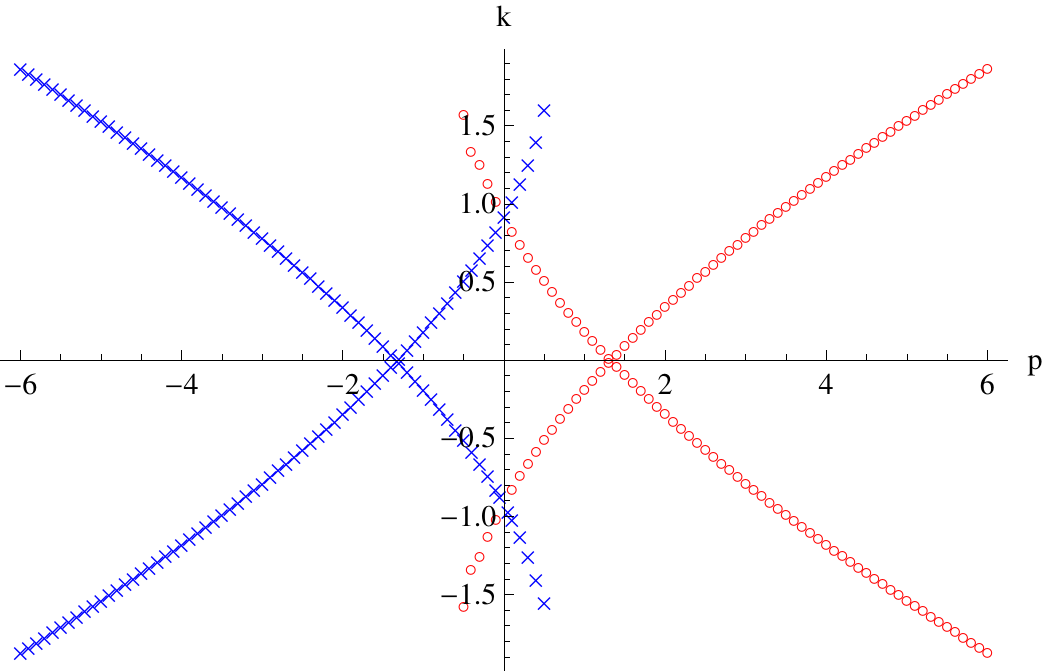}
\caption{Poles at $k=k_F$ (blue lines) and zeroes at $k=k_L$ (red lines) \emph{vs.}~$p$ with $q=1$. Notice the symmetry under $k\to -k$, and the duality of poles and zeroes under $p\to -p$.}
\label{figpoleszeroes}
\end{figure}

Finally, figure \ref{figpoleszeroes} shows the location of poles
($k=k_F$)  and zeroes ($k=k_L$) as the dipole coupling strength
$p$ varies. Notice the symmetry under $k\to -k$, as well as the
interchange $k_L \leftrightarrow k_F$ under the mapping $p\to -p$.

\section{Conclusions and discussion}
\label{sect5}

 We have shown that a holographic theory with a bulk
dipole interaction between a massless fermion and gauge field
possesses a robust phase diagram including Fermi and non-Fermi
liquids, insulating Mott state and pseudo-gap state. The various
phases are identified by the structure of the poles and zeroes found in $G_R$.
The selection of the phase is controlled by the strength of the
dipole coupling which
plays the role of an order parameter in the
holographic system \cite{Edalati:2010ww,Edalati:2010ge}.

  A pole of $G_R$ is indicative of a Fermi  or non-Fermi fluid
while a zero is responsible for an insulating phase. It is the
coexistence of both that underlies the identification of a holographic
pseudo-gap
state. We showed that a duality exists relating systems of
opposite dipole coupling strength $p$. This duality maps zeroes to
poles and vice versa, pointing to the interesting possibility of
understanding the properties of a system with zeroes (insulating
phase) by mapping the system to one with poles (Fermi liquid).

It will be interesting to explore further the pseudo-gap state and
other aspects  that may be recovered by the evolution of the Fermi
and Luttinger surfaces.  Of particular significance will be the
response of the system with $\vec k$ dependence. Also, it might be
worth exploring further the physical significance of the behavior
of $\Im \det G_R$ near zeroes and poles.

As we discussed in the introduction, the 
interplay between the state of matter and the appearance of
poles and zeroes is also observed in QCD. The difference with QCD is
that the appearance of free quarks or bound states depends
on the dynamics of the strong gauge theory while in our case there
is a variation of a simple coupling that gives the various phases
of matter. It would be intriguing to find other dynamical systems
exhibiting this behavior.

One may wonder if this analysis can be applied to holographic
superconductors. In BCS superconductivity the role of the order
parameter is played by the condensate of a scalar field. The
formation of a condensate corresponds to a critical temperature
below which the system enters the superconductivity phase. In this
phase, the electrons are combined to form Cooper pairs which are
electronic bound states and therefore correspond to zeroes.
On
the contrary, the electrons in the normal phase are free
corresponding to poles.

In spite of the similarities with superconductivity, our
analysis does not explicitly break a gauge symmetry that
could have defined an order parameter for the various
phases. Instead the metallic
phases  are controlled by  an explicit coupling of
charged fermions to a gauge field.  Nevertheless, it will be
enlightening to consider how a bulk scalar responsible for
superconductivity or the coupling will influence the system. 
Also worth exploring is if a description with additional fields could
explain the Green function zeroes in terms of a composite excitation.

 Another way to better understand our results is to
calculate the holographic entanglement entropy in our theory. The
entanglement entropy \cite{Ryu:2006bv,Ryu:2006ef} has proven to be
a powerful tool in counting the degrees of freedom available in a
holographic system. In \cite{Kuang:2014kha} it was found that the
holographic entanglement entropy of a superconducting phase is
less than that of the normal phase due to the
 formation of Cooper pairs reducing the degrees of
freedom available (see also \cite{Albash:2012pd}). Near the
contact interface,  the normal phase entanglement entropy
possesses a higher value compared to the superconducting phase due
to the proximity effect: the leakage of Cooper pairs to the normal
phase so more free electrons were available. Similarly we expect
that the  entanglement entropy in a holographic fermionic system
with a dipole coupling will vary, as the degrees of freedom change
from the Fermi liquid phase to the Mott insulating phase providing
important information on the various phases of the system.
However, such a study has to overcome the basic difficulty of
working with a probe fermonic system. The boundary theory will see
 the effect of the dipole coupling only with a fully back-reacted solution. 
 Once a full solution is obtained the dipole 
coupling could leave some trace
on the metric. Work in this  direction is in progress.

\acknowledgments{We thank P.\ W.\ Phillips for illuminating discussions.
 E.P is partially supported by the Greek Ministry of Education and Religious Affairs, Sport and Culture through the ARISTEIA II action of the operational program Education and Lifelong Learning.}

\end{document}